\documentclass[12pt,preprint]{aastex}
\usepackage{emulateapj5} 

\slugcomment{Submitted to ApJL}

\shorttitle{GALEX observations of M32}
\shortauthors{Gil de Paz et al.}

\begin{document}

\title{GALEX observations of the UV surface brightness and color profiles of the Local Group elliptical galaxy M32 (NGC221)}

\author{Armando Gil de Paz\altaffilmark{1}, Barry F. Madore\altaffilmark{1},
Young-Jong Sohn\altaffilmark{2}, Young-Wook Lee\altaffilmark{2},
Mark Seibert\altaffilmark{3}, R. Michael Rich\altaffilmark{4},
Luciana Bianchi\altaffilmark{5}, Tom A. Barlow\altaffilmark{3},
Young-Ik Byun\altaffilmark{2}, Jos\'{e} Donas\altaffilmark{6},
Karl Forster\altaffilmark{3}, Peter G. Friedman\altaffilmark{3},
Timothy M. Heckman\altaffilmark{7}, Patrick Jelinsky\altaffilmark{8},
Roger F. Malina\altaffilmark{6}, D. Christopher Martin\altaffilmark{3},
Bruno Milliard\altaffilmark{6}, Patrick Morrissey\altaffilmark{3},
Susan G. Neff\altaffilmark{9}, David Schiminovich\altaffilmark{3},
Oswald H. W. Siegmund\altaffilmark{8}, Todd Small\altaffilmark{3},
Alex S. Szalay\altaffilmark{7}, Barry Y. Welsh\altaffilmark{8},
Ted K. Wyder\altaffilmark{3}}

\altaffiltext{1}{Observatories of the Carnegie Institution of Washington, 813 Santa Barbara St., Pasadena, CA 91101; agpaz, barry@ociw.edu}

\altaffiltext{2}{Center for Space Astrophysics, Yonsei University, Seoul 120-749, Korea; sohnyj, ywlee, byun@csa.yonsei.ac.kr}

\altaffiltext{3}{California Institute of Technology, MC 405-47, 1200 East California Boulevard, Pasadena, CA 91125; mseibert, tab, krl, friedman, cmartin, patrick, ds, tas, wyder@srl.caltech.edu}

\altaffiltext{4}{Department of Physics and Astronomy, University of California, Los Angeles, CA 90095; rmr@astro.ucla.edu}

\altaffiltext{5}{Center for Astrophysical Sciences, The Johns Hopkins University, 3400 N. Charles St., Baltimore, MD 21218; bianchi@skysrv.pha.jhu.edu}

\altaffiltext{6}{Laboratoire d'Astrophysique de Marseille, BP 8, Traverse du Siphon, 13376 Marseille Cedex 12, France; jose.donas, roger.malina, bruno.milliard@astrsp-mrs.fr}

\altaffiltext{7}{Department of Physics and Astronomy, The Johns Hopkins University, Homewood Campus, Baltimore, MD 21218; heckman, szalay@pha.jhu.edu}

\altaffiltext{8}{Space Sciences Laboratory, University of California at Berkeley, 601 Campbell Hall, Berkeley, CA 94720; patj, ossy, bwelsh@ssl.berkeley.edu}

\altaffiltext{9}{Laboratory for Astronomy and Solar Physics, NASA Goddard Space Flight Center, Greenbelt, MD 20771; neff@stars.gsfc.nasa.gov}

\begin{abstract}
M32, the compact elliptical-galaxy companion to the Andromeda spiral
galaxy has been imaged by the Galaxy Evolution Explorer (GALEX) in two
ultraviolet bands, centered at $\sim$1500 (FUV) and 2300\,\AA\
(NUV). The imaging data have been carefully decomposed so as to
properly account for the complicated background contamination from the
disk of M31. We have derived the surface brightness and color profiles
finding a slightly positive color gradient of
$\Delta$($FUV$--$B$)/$\Delta$$\log$\,(r)=$+$0.15$\pm$0.03 within one
effective radius. Earlier data from the Ultraviolet Imaging Telescope
suggested that M32 had an extremely large (negative) FUV-optical color
gradient ($\Delta$($FUV$--$B$)/$\Delta$$\log$\,(r)$<$$-$2), inverted with
respect to the majority of gradients seen in giant elliptical
galaxies. Our new results show that, despite of its very low
UV-upturn, M32 has similar UV properties to those observed in luminous
elliptical galaxies.
\end{abstract}


\section{Introduction}
\label{intro}
The compact elliptical galaxy M32 has been widely used in the past as
template for the study of stellar populations and chemical evolution
of elliptical galaxies (e.g$.$ Freedman 1992 and references
therein). It is very nearby (780\,kpc; Tonry 1991; Freedman \& Madore
1990). It has very high surface brightness at optical wavelengths and
it is of high metallicity ($-$0.2$<$[Fe/H]$<$$+$0.01; Grillmair et
al$.$ 1996).

However, the initial study of the UV properties of this galaxy by
O'Connell et al$.$ (1992) and more recently by Ohl et al$.$ (1998)
cast some doubt on M32 as being a truly typical example of an
elliptical galaxy. These authors, using data from the Shuttle-borne
Ultraviolet Imaging Telescope (UIT), claimed the presence of a strong
FUV-optical color gradient in M32, but inverted with respect to the
gradients observed in the vast majority of elliptical galaxies. While
in regular, luminous elliptical galaxies the inner regions are
slightly bluer than the outer parts (probably suggesting a stronger
UV-upturn in the nuclear regions), for M32 these authors reported the
opposite: a very strong blue trend ($\sim$3\,mag within the effective
radius; $r_{\mathrm{eff}}$) toward outer regions of the galaxy.

Newly obtained GALEX FUV observations now show that the FUV-optical
gradient in M32 ($\Delta$($FUV$--$B$)/$\Delta$$\log$\,(r)=$+$0.15$\pm$0.03) is
in fact very similar to the gradients commonly measured in luminous
elliptical galaxies. This analysis rests on a careful subtraction of
the background emission from the disk of M31 (see e.g$.$ Choi,
Guhathakurta, \& Johnston 2002). We suggest that the strong negative
gradient reported by Ohl et al$.$ (1998) may have been caused by
problems in the density-to-flux calibration of UIT photographic data
at low surface-brightness levels.

\section{Observations}
\label{observations}
GALEX has recently completed a mosaic image of the entire Andromeda
galaxy. This mosaic includes observations of the compact elliptical
galaxy M32 with exposure times of 6138 seconds in the FUV band
($\lambda$=1530\,\AA) and 4808 seconds in the NUV band
($\lambda$=2310\,\AA). The final spatial resolution (FWHM) of the
combined images of M32 used in this Letter was 6.0\arcsec\ and
6.8\arcsec\ for the FUV and for the NUV. The images were flux
calibrated using the GALEX zero points (Morrissey et al$.$ 2004).

In Figures~\ref{fig1}a \& \ref{fig1}b we show a
25\arcmin$\times$25\arcmin\ section of the GALEX FUV and NUV images
centered on M32. It is evident from these figures that significant FUV
and NUV emission from the disk of M31 seriously affects M32 and that
it is complex with a steep NW-SE gradient. The average FUV (NUV)
background associated with the disk of M31 that we measure close to
the position of M32 is 26.0 (25.7) mag\,arcsec$^{-2}$, while the
background observed far from the disk of M31 is much lower, 27.2
(26.7) mag\,arcsec$^{-2}$. Therefore, if we want to derive reliable
surface photometry for M32, detailed modeling of the M31 disk emission
is required.

Finally, we complemented our GALEX observations with archival $HST$
data obtained with the STIS FUV MAMA (G0 9053; PI: T.M$.$ Brown). The
$HST$ image allows us to analyze the innermost 16\arcsec\ (in radius)
of M32 at high spatial resolution ($<$0.1\arcsec).

\section{Analysis}
\label{analysis}

\subsection{Subtraction of the Disk of M31}

\begin{figure}
\epsscale{0.66}
\plotone{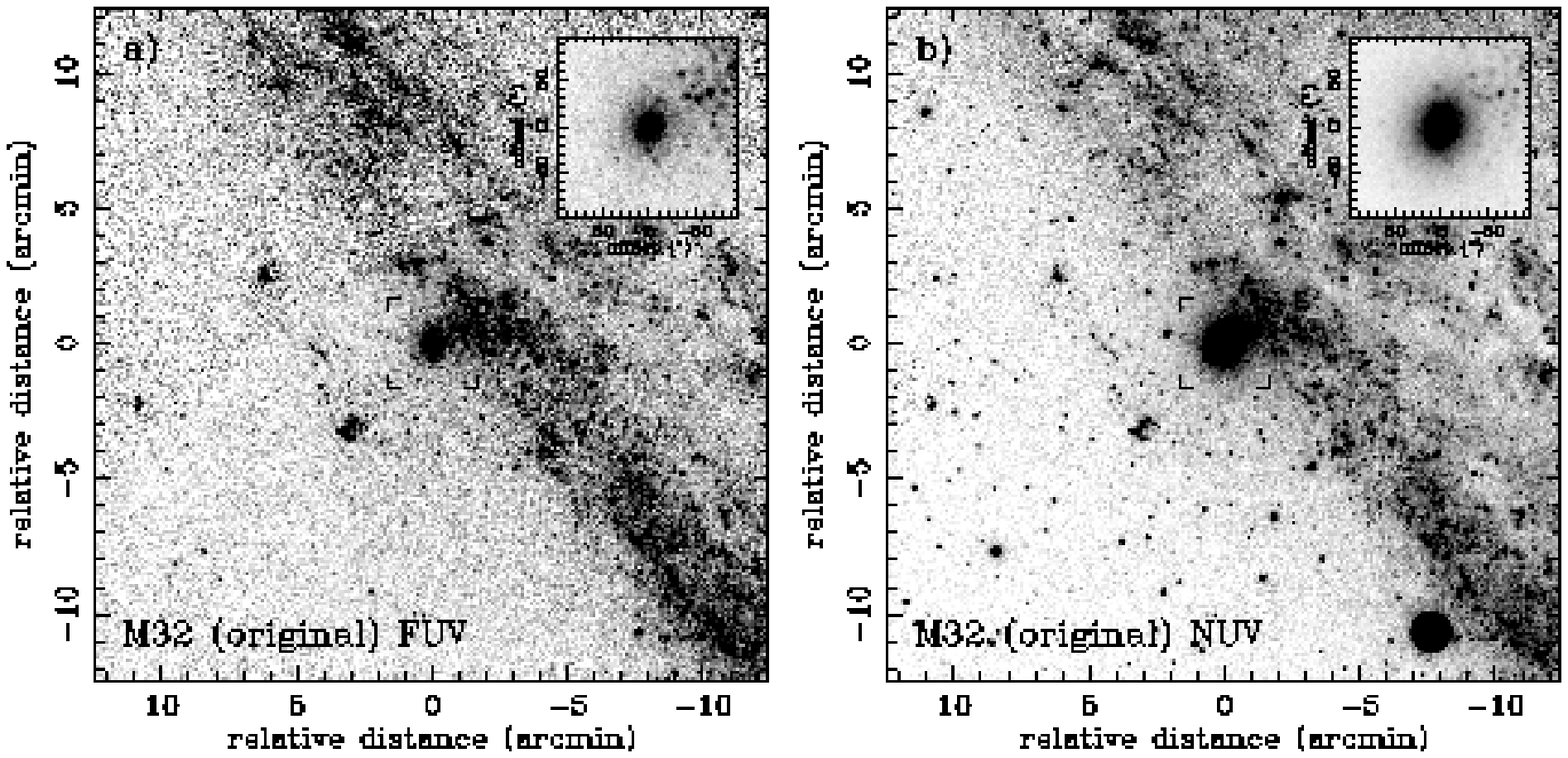}
\epsscale{0.66}
\plotone{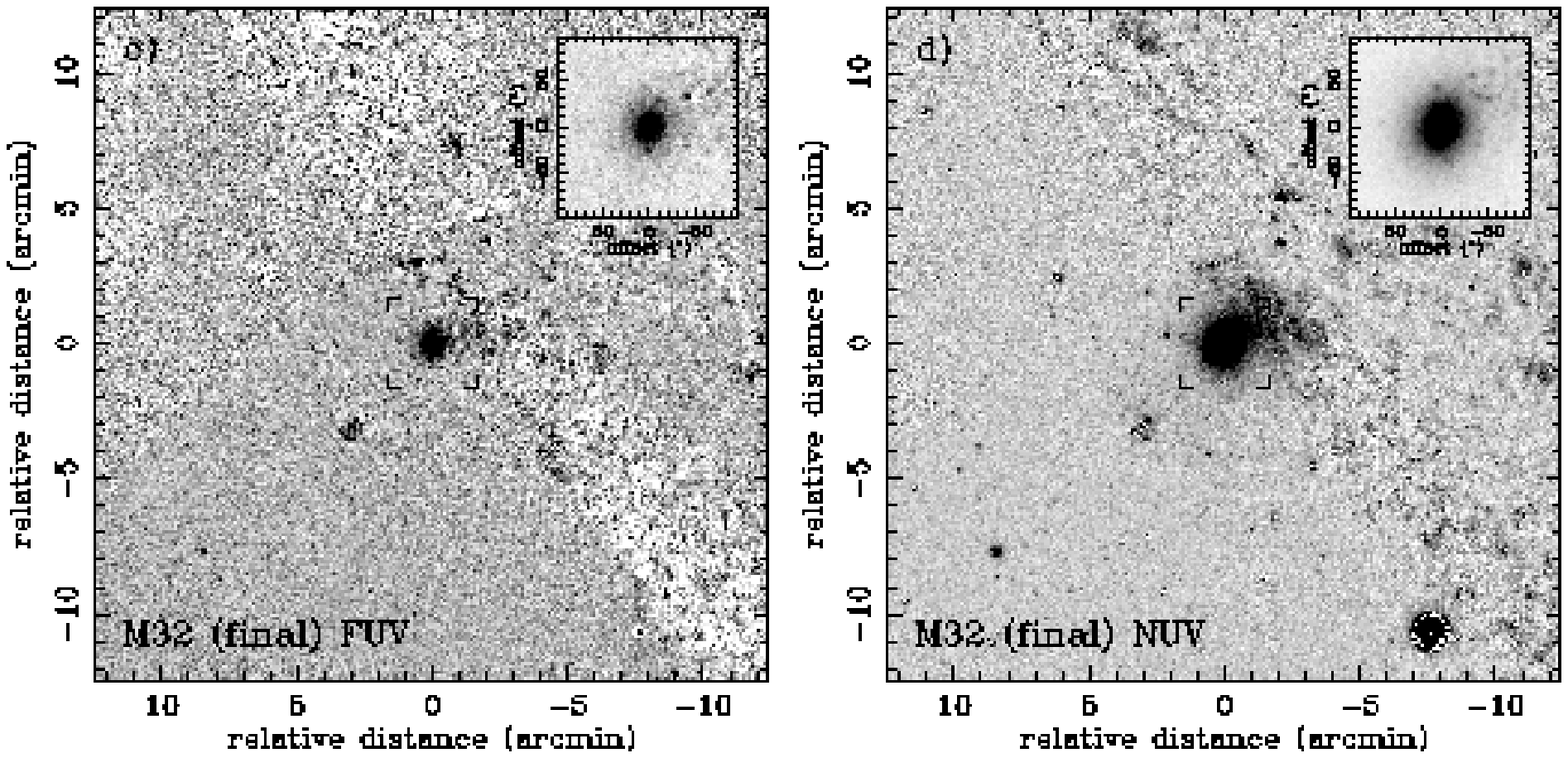}
\caption{GALEX UV images of M32. {\bf a)} Combined GALEX FUV image corresponding to a total exposure time of 6138 seconds at the position of M32. A 2$\times$ expanded view centered on M32 is shown in the upper-right corner of each plot (with a different stretch). {\bf b)} The same as {\bf a} for the NUV channel. The total exposure time at the position of M32 for this image was 4808 seconds.  {\bf c)} GALEX FUV image after the subtraction of point sources and the unresolved background from M31. {\bf d)} The same as {\bf c} for the NUV channel.\label{fig1}}
\end{figure}

The morphology of the disk of M31 both in the FUV and NUV is very
clumpy (see Figures~\ref{fig1}a \& \ref{fig1}b), mostly due to the
distinct contribution of OB associations and HII regions. This makes
the modelling of the disk more complicated than at optical wavelengths
where the light distribution is significantly smoother and can be
reasonably well reproduced by an exponential disk (Peletier 1993; Choi
et al$.$ 2002).

The subtraction of the disk of M31 was carried out in two
steps. First, we removed the unresolved, diffuse background
component. For the purpose of modeling this background component we
masked all the individual clusters, associations, field stars, and M32
itself. Then, we divided the image into boxes of
75\arcsec$\times$75\arcsec\ and fitted a low-order polynomial to the
remaining (un-masked) pixels using the IRAF task {\sc surfit}. We then
subtracted the fitted background from the images and added the mean
value of the modelled sky.

In the second step of subtracting the M31 disk, we removed the
point-sources contribution by modelling the PSF of the GALEX images
using the IRAF task {\sc psf}. We then subtracted the point sources
previously identified by {\sc daofind} using the task {\sc substar}.

The final result from the subtraction of both the unresolved
background and point sources is shown in Figures~\ref{fig1}c \&
\ref{fig1}d for the FUV and NUV images, respectively. A few point
sources in the outer regions of the halo of M32 and residuals from the
point-source subtraction were further masked in order to derive the
surface brightness and color profiles of M32. 


\subsection{Surface Brightness and Color Profiles}

To compute the FUV and NUV surface brightness profiles of M32 we used
isophotal parameters derived by both Peletier (1993) and Choi et al$.$
(2002) at optical wavelengths. This allowed us to directly compare our
UV surface photometry with that derived in the optical and obtain
self-consistent UV-optical color profiles. Note that in both of the
above papers the contamination from the disk of M31 was explicitly
accounted for. For sake of comparison we also fitted isophotes to our
final NUV image using the iterative method of Jedrzejewski (1987). We
found very small differences both in ellipticity ($-$0.04$\pm$0.04)
and position angle (2$\pm$6$^{\circ}$) between our best-fitting
isophotes and those of Peletier (1993).
 
\begin{figure}
\epsscale{0.45}
\plotone{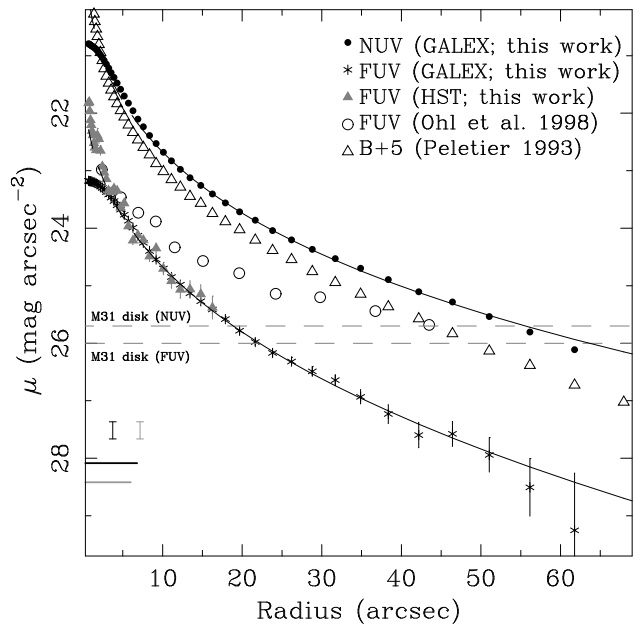}
\caption{Surface brightness profiles of M32. The surface brightness
profiles of M32 in the GALEX FUV, NUV, and optical $B$ bands are shown
(we have adopted the isophotal parameters derived by Peletier
1993). The best-fitting S\'ersic laws for the FUV and NUV profiles are
also shown. The FUV surface photometry published by Ohl et al. (1998)
is shown as open circles in this plot. Grey triangles indicate the FUV
surface brightness profile obtained from archival $HST$-STIS data. The
absolute calibration errors of 0.15\,mag and the PSF FWHM of the FUV
and NUV images are shown in the lower-left corner of the plot using
grey and black tick marks, respectively. The horizontal broken lines
are the approximate levels of the M31 background contamination at the
position of M32 in the FUV and NUV bands.\label{fig2}}
\end{figure}

In Figure~\ref{fig2} we show the FUV and NUV surface brightness
profiles (in AB magnitudes) obtained from GALEX observations. The
equivalent isophotal radius in this plot is computed as $\sqrt{a
\times b}$. The $B$-band and UIT FUV surface brightness profiles
published by Peletier (1993) and Ohl et al$.$ (1998), respectively,
are also plotted for comparison.


The best-fitting S\'ersic-law indices of the FUV and NUV surface
brightness profiles shown in Figure~\ref{fig2} are 0.38$\pm$0.01 and
0.26$\pm$0.01, respectively. These values are very similar to what is
expected for a pure de Vaucouleurs profile (0.25). Note that our UV
observations do not reach the larger galactocentric distances where
Graham (2002) reported the presence of an extended exponential disk.

For the innermost regions of the FUV surface brightness profile of M32
we have used an archival $HST$-STIS image. The background was
estimated by matching the outermost part of the $HST$ profile
(obtained using ground-based optical isophotal parameters) with the
GALEX FUV profile (see Figure~\ref{fig2}).

As seen in Figure~\ref{fig3}a all the color gradients obtained are
rather flat within an $r_{\mathrm{eff}}$ (32\arcsec; e.g$.$ Choi et
al$.$ 2002). Optical data used in this plot come from Peletier (1993);
however almost identical results are obtained if data from Choi et
al$.$ (2002) are used. It is noteworthy that this behavior is observed
even at distances as close to the galaxy center as 2\arcsec, below
which atmospheric seeing starts to affect ground-based optical
photometry (see Figure~\ref{fig3}b).

\section{Discussion}
\label{discussion}

\subsection{M32 and the Origin of the UV-upturn}

\begin{figure}
\epsscale{0.85}
\plotone{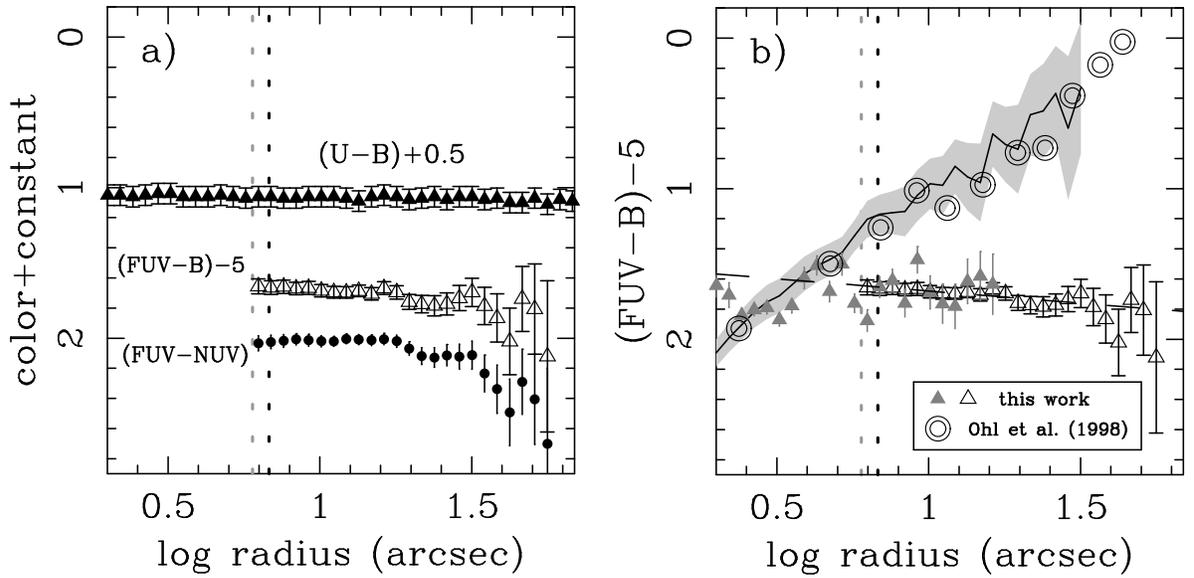}
\caption{Color profiles of M32. {\bf a)} ($U$$-$$B$), ($FUV$--$B$), and
($FUV$--$NUV$) color profiles of M32. Optical data have been taken from
Peletier (1993). Vertical grey and black broken lines show the
position of the PSF FWHM of the FUV and NUV images, respectively. {\bf
b)} ($FUV$--$B$) color profile obtained by GALEX (open triangles)
compared with that obtained using the published FUV photometry data of
Ohl et al$.$ (1998) (circled circles). The black solid curve and the
grey-shaded areas show the mean and $\pm$1-$\sigma$ color profile
obtained by us using the same background subtraction procedure for the
GALEX data described in the body of this Letter, but applied to the
UIT image. The best-fitting ($FUV$--$B$) color gradient to the GALEX
data (not including the $HST$-STIS data) is also shown as a broken
line. Grey triangles show the color profile of the innermost
16\arcsec\ of M32 obtained from $HST$-STIS FUV and ground-based
$B$-band photometry.\label{fig3}}
\end{figure}

A least-squares fit to the ($FUV$--$B$) color profile between
6\arcsec\ (FWHM) and 32\arcsec\ ($r_{\mathrm{eff}}$) in radius yields
a color gradient of
$\Delta$($FUV$--$B$)/$\Delta$$\log$\,(r)=$+$0.15$\pm$0.03 (see
Figure~\ref{fig3}b). This value is similar to that obtained by Ohl et
al$.$ (1998) for luminous elliptical galaxies ($+$0.5$\pm$0.3), but it
is very different from that obtained for M32 by Ohl et al$.$ (1998)
($\Delta$($FUV$--$B$)/$\Delta$$\log$\,(r)$<$$-$2). First, we checked
whether the difference found arises from the UIT surface photometry
obtained by Ohl et al$.$ (1998) being significantly affected by the
emission of the disk of M31. To check this we derived the same
($FUV$--$B$) color gradient using the background subtraction procedure
described above on the archival Astro-1 (B1 filter;
$\lambda_{\mathrm{eff}}$=1520\,\AA) FUV UIT image. The color profile
obtained is remarkably similar to that obtained by Ohl et al$.$ (1998)
(see Figure~\ref{fig3}b) after being offset to match the Astro-2 (B5
filter; $\lambda_{\mathrm{eff}}$=1615\,\AA) FUV UIT photometry. We
also studied the effects of the wings of the UIT PSF on the
($FUV$--$B$) profile. The maximum impact of this effect on the
($FUV$--$B$) color gradient is found to be $\leq$0.4\,mag within the
central 30\arcsec. The other possibility is that this difference may
be a problem in the density-to-flux calibration of the (photographic)
UIT image at very low surface-brightness levels. However, a detailed
study of the linearity of the UIT data is beyond the scope of this
Letter.

The GALEX ($FUV$--$NUV$) color gradient (also sensitive to the strength of
the UV-upturn) is similar to that derived for ($FUV$--$B$). This
confirms that the color gradient derived is real and not an artifact
introduced by the different background-subtraction technique or
because of spatial resolution differences between the UV and optical
data. This is confirmed by the analysis of the archival $HST$-STIS
data, which shows a gradient at the innermost regions of the galaxy
similar to and extending that obtained from GALEX observations alone
(see Figure~\ref{fig3}b).

Despite its very shallow UV-upturn, which in principle could be
explained by emission from post-AGB stars, Brown et al$.$ (2000) have
shown that the UV emission in M32 is dominated by hot HB stars. By
analogy, this suggests that hot HB stars are also responsible for most
of the FUV emission associated with the UV-upturn observed in luminous
elliptical galaxies (see Brown et al$.$ 1997). The ($FUV$--$B$) color
gradient reported in this Letter, similar to that measured in luminous
elliptical galaxies, along with the results of Brown et al$.$ (2000)
suggest that the properties and spatial distribution of the hot HB
stars in M32 are the same as those in luminous elliptical
galaxies. The great advantage here being that M32 is the only object
where individual hot-HB stars have actually been resolved.


Burstein et al$.$ (1988) claimed that elliptical galaxies with larger
Mg$_2$ indices show stronger UV-upturns. This has been interpreted as
resulting from a dependence of mass-loss efficiency and helium
abundance on metallicity (Greggio \& Renzini 1990; O'Connell
1999). However, Rich et al$.$ (2004), using a large sample of
low-redshift red galaxies, do not find any correlation between the
strength of the UV-upturn and the Mg$_2$, D4000, H$\beta$ indices or
the velocity dispersion (see also Deharveng, Boselli, \& Donas
2002). Ohl et al$.$ (1998) also reported a lack of correlation between
the ($FUV$--$B$) gradient and the Mg$_2$-index gradient. These results
suggest the presence of a second parameter (decoupling from the
Fe-peak, helium abundance, age; O'Connell 1999) that could have an
impact even stronger than the metallicity on the evolution of the
UV-upturn.

The suggested presence of a strong negative gradient in ($FUV$--$B$)
color in M32 (Ohl et al$.$ 1998), where the existence of an
intermediate age stellar population (spatially segregated toward the
galaxy center) has been frequently proposed (Grillmair et al$.$ 1996),
had been claimed as an indication that the age may play a significant
role in the evolution of the UV-upturn. However, our results in
combination with the lack of structure in the optical-colors and
spectroscopic-index maps of M32 (e.g$.$ del Burgo et al$.$ 2001)
indicate that if this stellar population is present, it is very
smoothly distributed across the body of the galaxy and that the
properties of the hot HB responsible for the UV-upturn are also very
similar at any position in the galaxy.

\subsection{Is M32 a Peculiar Object?}

The two most intriguing differences reported between the properties of
M32 and those of luminous elliptical galaxies had been: (1) the
presence of an intemediate-aged stellar population (e.g$.$ Grillmair
et al$.$ 1996) and (2) the large (inverted) ($FUV$--$B$) color measured
by Ohl et al$.$ (1998). Regarding the former problem we note that many
of the spectral synthesis analyses carried out to date assume a pure
red clump HB morphology, while Brown et al$.$ (2000) have identifed a
large population of hot HB stars in M32. With regard to the latter
topic, our results show that the previously reported unusual
($FUV$--$B$) color gradient does not exist and that the UV properties of
M32 are very similar to those of luminous ellipticals.

We conclude that, although M32 is certainly an extreme example in the
sequence of elliptical galaxies in many of its properties and the
possible presence of an intermediate-aged stellar population should
not be ignored, it cannot be considered to be a peculiar object and
its use as a reference object for stellar populations synthesis is
justified.


In summary, the analysis of GALEX FUV and NUV imaging data of the
compact elliptical galaxy M32 yields very small (positive) ($FUV$--$B$)
and ($FUV$--$NUV$) color gradients, comparable to values seen in luminous
elliptical galaxies. This result suggests that the properties of the
hot HB stars responsible for the formation of the (very weak)
UV-upturn in M32 are not a strong function of position in the galaxy
and that they are probably similar to hot HB stars in luminous
elliptical galaxies.

\acknowledgments

GALEX (Galaxy Evolution Explorer) is a NASA Small Explorer, launched
in April 2003. We gratefully acknowledge NASA's support for
construction, operation, and science analysis for the GALEX mission,
developed in cooperation with the Centre National d'Etudes Spatiales
of France and the Korean Ministry of Science and Technology. We thank
Robert W$.$ O'Connell and Jean-Michel Deharveng for valuable comments.

\clearpage 

\end{document}